\documentclass[preprint,aps,onecolumn,superscriptaddress]{revtex4-1}

\usepackage{graphics}
\usepackage{epsfig}
\usepackage{multirow}
\usepackage{hhline}
\usepackage{cellspace}
\usepackage{caption}
\usepackage{amsmath,amssymb,amscd}

\begin{document}

\title {When do tripdoublet states fluoresce? A theoretical study of copper(II) porphyrin} 

\author{Xingwen Wang}
\affiliation{Qingdao Institute for Theoretical and Computational Sciences, Shandong University, Qingdao 266237, China}

\author{Chenyu Wu}
\affiliation{Qingdao Institute for Theoretical and Computational Sciences, Shandong University, Qingdao 266237, China}

\author{Zikuan Wang}\email{zwang@kofo.mpg.de}
\affiliation{Qingdao Institute for Theoretical and Computational Sciences, Shandong University, Qingdao 266237, China}
\affiliation{Max-Planck-Institut f\"ur Kohlenforschung, Kaiser-Wilhelm-Platz 1, Mülheim an der Ruhr 45470, Germany}

\author{Wenjian Liu}\email{liuwj@sdu.edu.cn}
\affiliation{Qingdao Institute for Theoretical and Computational Sciences, Shandong University, Qingdao 266237, China}


\begin{abstract}

Open-shell molecules rarely fluoresce, due to their typically faster non-radiative relaxation rates compared to closed-shell ones. Even rarer is the fluorescence from states that have two more unpaired electrons than the open-shell ground state, for example tripdoublet states (a triplet excitation antiferromagnetically coupled to a doublet state). The description of the latter states by U-TDDFT is notoriously inaccurate due to large spin contamination.
In this work, we applied our spin-adapted TDDFT method, X-TDDFT, and the static-dynamic-static second order perturbation theory (SDSPT2), to the study of the excited states as well as their relaxation pathways of copper(II) porphyrin; previous experimental works suggested that the photoluminescence of some substituted copper(II) porphyrins originate from a tripdoublet state, formed by a triplet ligand $\pi\to\pi^*$ excitation.
Our results demonstrated favorable agreement between the X-TDDFT, SDSPT2 and experimental excitation energies, and revealed noticeable improvements of X-TDDFT compared to U-TDDFT, suggesting that X-TDDFT is a reliable tool for the study of tripdoublet fluorescence.
Intriguingly, the aforementioned tripdoublet state is the lowest doublet excited state and lies only slightly higher than the lowest quartet state, which explains why the tripdoublet of copper(II) porphyrin is long-lived enough to fluoresce; an explanation for this unusual state ordering is given.
Indeed, thermal vibration correlation function (TVCF)-based calculations of internal conversion, intersystem crossing, and radiative transition rates confirm that copper(II) porphyrin emits thermally activated delayed fluorescence (TADF) and a small amount of phosphorescence at low temperature (83 K), in accordance with experiment. The present contribution is concluded by a few possible approaches of designing new molecules that fluoresce from tripdoublet states.

\end{abstract}

\maketitle

\section{Introduction}

\begin{figure}
	\centering
	\includegraphics[width=0.8\textwidth]{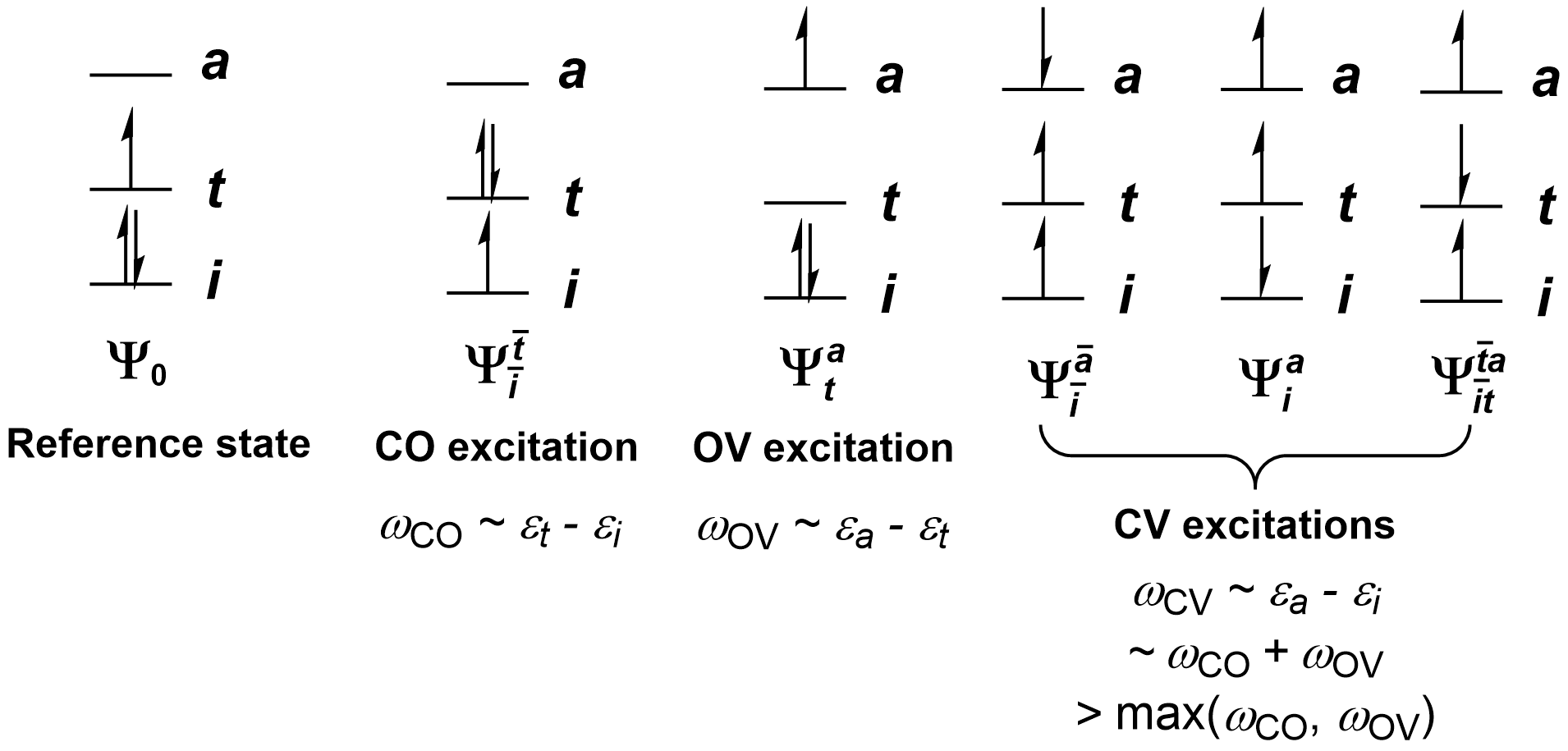}
	\caption{
		Schematic depictions of closed-open (CO), open-vacant (OV), and closed-vacant (CV) excitations, and their approximate excitation energies as predicted from restricted open-shell Kohn-Sham (ROKS) orbital energy differences.
	}
	\label{CO_OV_CV}
\end{figure}

Fluorescence, while ubiquitous in organic and organometallic molecules, is in most cases observed in closed-shell systems. It is well-known that introducing an open-shell impurity, such as dioxygen\cite{photosensitize_singlet_oxygen}, a stable organic radical\cite{nitroxyl_radical_quench} or a transition metal ion\cite{metal_flavine_quench}, frequently quenches the fluorescence of a closed-shell molecule\cite{Evans_radical_quench}. One reason of this phenomenon is that the addition of an unpaired electron to a system typically introduces additional low-lying states, in particular charge transfer states that involve an electron exciting from or out of the new open-shell orbital (O). Moreover, while spin-conserving single excitations of a singlet reference determinant from closed-shell (C) to vacant-shell (V) orbitals, hereafter termed CV excitations following our previous works\cite{STDDFT1,STDDFT2,XTDDFT,XTDDFTgrad}, give rise to $n_{\rm{C}}n_{\rm{V}}$ singlet excited states and $n_{\rm{C}}n_{\rm{V}}$ triplet excited states (where $n_{\rm{C}}$ and $n_{\rm{V}}$ are the number of closed-shell an vacant-shell orbitals, respectively), with an $M_S=1/2$ doublet determinant one obtains $2n_{\rm{C}}n_{\rm{V}}$ excitations that are mixtures of doublets and quartets (the $\Psi_{\bar{i}}^{\bar{a}}$ and $\Psi_{i}^{a}$ determinants in Figure~\ref{CO_OV_CV}; here orbitals without overbars denote $\alpha$ orbitals, and those with overbars denote $\beta$ ones). They can be linearly combined to make $n_{\rm{C}}n_{\rm{V}}$ pure doublet states, but the other linear combination remains a mixture of doublet and quartet:
\begin{eqnarray}
\Psi_{\rm{singdoublet}} & = & \frac{1}{\sqrt{2}} \left(\Psi_i^a + \Psi_{\bar{i}}^{\bar{a}}\right), \label{Psi_singdoublet} \\
\Psi_{\rm{mixed}} & = & \frac{1}{\sqrt{2}} \left(\Psi_i^a - \Psi_{\bar{i}}^{\bar{a}}\right). \label{Psi_mixed}
\end{eqnarray}
In spin-adapted TDDFT methods, the latter are spin-adapted to give $2n_{\rm{C}}n_{\rm{V}}$ pure doublet states and $n_{\rm{C}}n_{\rm{V}}$ quartet states, by mixing with the $n_{\rm{C}}n_{\rm{V}}$ spin flip-up excitations from the $M_S=-1/2$ component of the reference determinant, i.e.~the $\Psi_{\bar{i}t}^{\bar{t}a}$ determinants in Figure~\ref{CO_OV_CV}\cite{STDDFT1,STDDFT2,XTDDFT}:
\begin{eqnarray}
\Psi_{\rm{tripdoublet}} & = & \frac{1}{\sqrt{6}} \left(- \Psi_i^a + \Psi_{\bar{i}}^{\bar{a}} + 2 \Psi_{\bar{i}t}^{\bar{t}a}\right), \label{Psi_tripdoublet} \\
\Psi_{\rm{quartet}} & = & \frac{1}{\sqrt{3}} \left(\Psi_i^a - \Psi_{\bar{i}}^{\bar{a}} + \Psi_{\bar{i}t}^{\bar{t}a}\right). \label{Psi_quartet}
\end{eqnarray}
Note that both the ``singdoublets'' and ``tripdoublets'' are pure doublet states. While the singdoublets Eq.~\ref{Psi_singdoublet} (which we called the CV(0) states in our previous works\cite{STDDFT1,STDDFT2,XTDDFT}) are direct analogs of singlet excited states out of a singlet reference, the tripdoublets Eq.~\ref{Psi_tripdoublet} (CV(1) states) do not have analogs in closed-shell systems, and create extra spin-allowed non-radiative relaxation pathways compared to when the reference determinant is singlet. This further contributes to the short excited state lifetimes of doublet systems. As a consequence, doublet molecules (and open-shell molecules in general) are rarely fluorescent.
\\\indent
Still, there exist open-shell molecules that do fluoresce, which have found applications in e.g.~organic light-emitting diodes (OLEDs)\cite{suo_doublet_emission,FengLi_doublet_emission_review}. However, their fluorescence usually originates from an excited state that has only one unpaired electron, i.e.~a CO or OV excited state (where CO stands for a single excitation from a closed-shell orbital to an open-shell one; similar for OV), instead of a CV excited state.
This can be partly rationalized by approximating the excitation energies of the system by orbital energy differences. Under this approximation, there is at least one CO state and one OV state below any given CV state, since the lowest CV excitation energy is the sum of the excitation energies of a CO state and an OV state (Figure~\ref{CO_OV_CV}). Therefore, the lowest CV state tends to not be the lowest excited state of the system, and thus usually has more energetically accessible non-radiative relaxation pathways than the low-lying CO and OV states do, rendering fluorescence from CV states especially hard to achieve.
To counter this, one may try to inhibit the non-radiative relaxation of the CV state to lower excited states. However, the sheer number of non-radiative relaxation pathways that one would have to inhibit poses a great challenge for designing an open-shell molecule that fluoresces from a CV state. Alternatively, one may design a system where the orbital energy difference approximation fails dramatically, allowing the lowest CV state to become the first excited state. In this case, the fluorescence from the CV state only needs to compete with the intersystem crossings (ISCs) to the lowest quartet state(s) and the internal conversion (IC) to the ground state, which are the only two energy downhill non-radiative relaxation pathways available to the CV state. In particular, note that when the CV excitations shown in Figure~\ref{CO_OV_CV} linearly combine to give singdoublets, tripdoublets and quartets via Eqs.~\ref{Psi_tripdoublet}-\ref{Psi_quartet}, there is an energy splitting that usually places the quartet below the tripdoublet, and the tripdoublet below the singdoublet; while the former is a consequence of Hund's rule, the latter can be rationalized by applying Hund's rule after neglecting the coupling of the open-shell orbital to the closed-shell and vacant-shell ones. This gives tripdoublets a much greater chance than singdoublets for emitting fluorescence with an appreciable quantum yield. Nevertheless, the singdoublet-tripdoublet splitting appears to be small in general, compared to the orbital energy difference that one would have to overcome, which can amount to several eVs. Hence, even the fluorescence from tripdoublets proves to be scarce.
\\\indent
The present paper represents a preliminary attempt to unveil some of the factors that enable an open-shell molecule to fluoresce from a tripdoublet state, via a case study of copper(II) porphyrin complexes. Copper(II) porphyrin complexes, like most porphyrin complexes, show two intense visible absorption bands near 390-420 nm and 520-580 nm\cite{Gouterman1959study,eastwood1969porphyrins}; they are conventionally termed the B and Q bands, respectively. Gouterman et al.\cite{eastwood1969porphyrins}~studied the luminescence of copper(II) porphyrin molecules in the solid state by exciting their Q bands, suggesting that the emission may originate from one of the two low-lying $\pi\to\pi^*$ states, $^{2}$T or $^{4}$T (here the 2, 4 represent the overall spin multiplicity of the complex, and T denotes that the ``local'' spin multiplicity of the porphyrin ring is triplet). They speculated that a rapid equilibrium may exist between the $^2$T and $^4$T states. The equilibrium ratio of these two states is largely dependent on the energy gap ($\Delta E_{\rm{DQ}}$) between them and the temperature, via the Boltzmann distribution. The radiative transition from the $^2$T state to the ground state is spin-allowed, making it much faster than the phosphorescence from the $^4$T state. Thus, when $\Delta E_{\rm{DQ}}$ is small and the temperature is high, the experimentally observed rapid emission is predominantly from the $^2$T state. Conversely, when $\Delta E_{\rm{DQ}}$ is large and the temperature is low, a slow emission attributed to the phosphorescence of the $^4$T state was observed instead, due to the concentration of the $^4$T state largely overwhelming that of the $^2$T state. Thus, molecules such as copper 2,3,7,8,12,13,17,18-octaalkylporphyrin (CuOAP), which possess small $\Delta E_{\rm{DQ}}$ values, exhibit luminescence primarily in the form of fluorescence from the $^2$T state at liquid nitrogen temperature, whereas copper 5,10,15,20-tetraphenylporphyrin (CuTPP) with a larger $\Delta E_{\rm{DQ}}$ mainly undergoes phosphorescence from the $^4$T state at the same temperatures. The unsubstituted copper porphyrin (CuP) is the most interesting of all, as pure phosphorescence was observed at low temperatures (35 K), which gradually gives way to fluorescence when the temperature was elevated, eventually giving pure fluorescence at 143 K\cite{CuP_hightemp_lumin}. Similar results have been obtained by following works with different techniques and/or solvents\cite{Kobayashi_CuP_AgP,Gouterman_SnPdCu}.
\\\indent
The simple and intuitive picture has since been supplemented by subsequent works, which also excited the B band, and proposed that charge transfer (CT) states may play an important role in the relaxation of the initial bright state to the essentially dark $^{2}$T state. Holten et al.\cite{yan1988effects} investigated the excited state relaxation processes of CuTPP and CuOEP at different temperatures and in different solvents, proposing possible pathways involving intermediate states that are probably ligand-to-metal CT (LMCT) states. This is supported by the gas-phase mass spectrometry experiments by Ha-Thi et al.\cite{ha2013efficient}, although the precise composition of the CT state remains uncertain. Understanding the excited-state relaxation pathways of copper porphyrins is crucial for gaining insights into their photophysical processes and controlling their optical properties. In particular, whether the CT state(s) (or any other excited states) lie below the $^{2}$T state may have a profound influence on whether the $^{2}$T state fluoresces or not, as follows from Kasha's rule. Meanwhile, the energy gap of the $^{2}$T and $^{4}$T states is important for the relative concentration of the two states, and therefore the relative intensities of fluorescence from the $^{2}$T state and the phosphorescence from the $^{4}$T state, i.e.~whether the experimentally observed luminescence should be attributed to fluorescence or phosphorescence, or both.
\\\indent
\begin{figure}[htbp]
	\centering
	\includegraphics[width=0.8\textwidth]{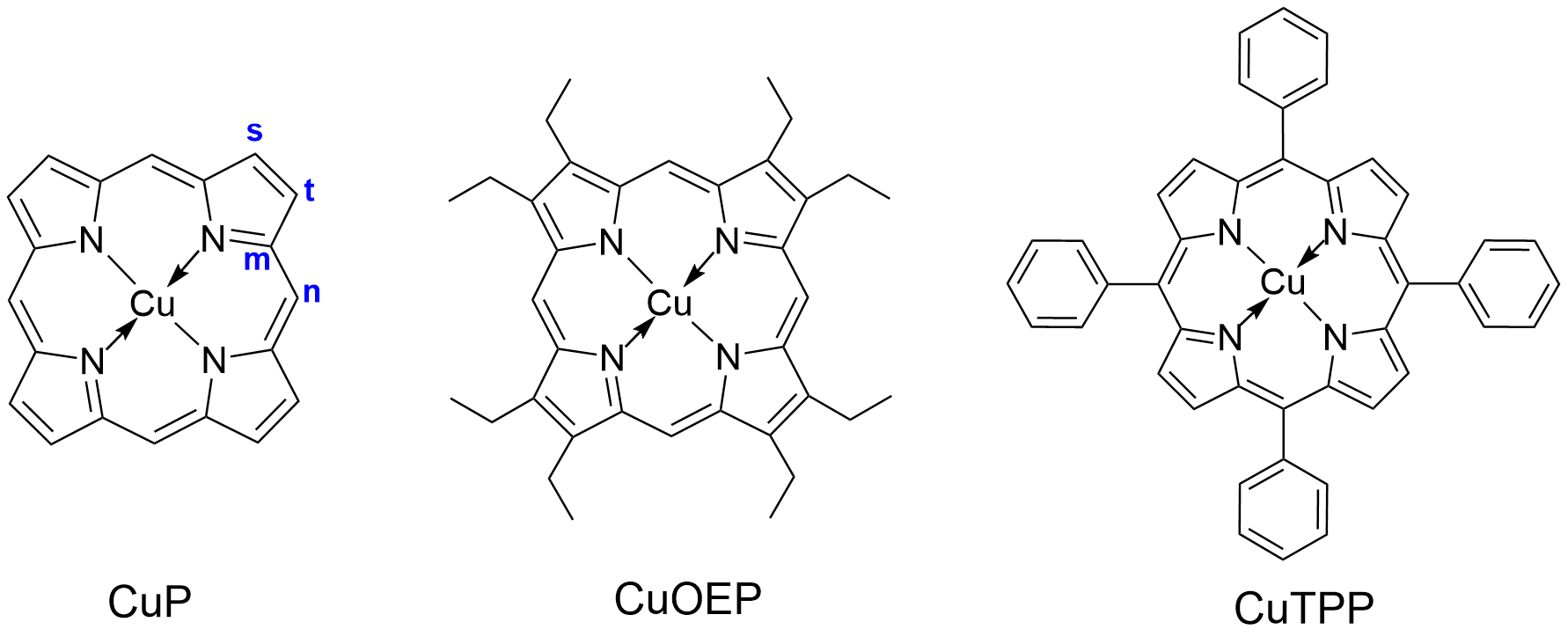}
	\caption{Molecular structures of CuP, CuOEP and CuTPP.}
	\label{structures}
\end{figure}
Despite the importance of tripdoublet fluorescence and the long history of experimental studies of copper porphyrins, accurate computational studies of this system prove to be difficult, as traditional unrestricted single-reference methods like U-TDDFT suffer from severe spin contamination issues, leading to systematically underestimated excitation energies. In particular, tripdoublet states are the worst scenario for U-TDDFT, as the errors of the U-TDDFT $\langle S^2 \rangle$ values of tripdoublet states reach the theoretical maximum of singly excited states, i.e.~2, when the reference state itself is not spin-contaminated\cite{STDDFT1,STDDFT2,XTDDFT,XTDDFTgrad,XTDDFTbenchmarkDD}. While multireference methods trivially solve the spin contamination problems, it is notoriously difficult to obtain an accurate multireference description of the electronic structure of metalloporphyrins, due to the complex interplay between static and dynamic correlation. In this study, we employed the methods developed by our group, namely X-TDDFT\cite{XTDDFT,XTDDFTgrad} and SDSPT2\cite{SDS,SDSPT2} (static-dynamic-static second-order perturbation theory), to address these challenges and provide a rational description of the photophysical processes in copper porphyrin molecules. As the first rigorous spin-adapted TDDFT method\cite{XTDDFT}, X-TDDFT gives spin-adapted excited states even when the reference state is open-shell, thereby generally giving better excitation energies, as well as better transition matrix elements involving the excited states. The recent development of the analytic gradient of X-TDDFT\cite{XTDDFTgrad} allowed us to use X-TDDFT for excited state geometry optimization and seminumerical Hessian calculations as well. For vertical excitation calculations, we could afford to use SDSPT2, which also served as a reference for benchmarking X-TDDFT and U-TDDFT.

\section{Computational details}

All DFT, TDDFT, and SDSPT2 calculations were performed using a development version of the Beijing Density Functional (BDF) package\cite{BDF1,BDF2,BDF3,BDF4,BDF5}. Geometry optimizations were conducted using the PBE0\cite{PBE01-1999,PBE02-1999} functional and x2c-SVPall\cite{x2c-svpall} basis set in the gas phase, including Grimme’s D3 dispersion correction\cite{DFT-D3-grimme2010effect,DFT-D3-grimme2011effect}, as implemented in the BDF software; relativistic effects were considered at the spin-free exact two component (sf-X2C) level\cite{X2C2009,X2C-2016,sf-X2C-2018,sf-X2C-2020}. For transition metal complexes (especially when excited states are considered), the choice of the optimum functional may not be obvious. Herein, four different functionals (BP86\cite{BP861-1988,BP862-1986,BP863-1986}, B3LYP\cite{B3LYP1,B3LYP2}, PBE0 and $\omega$B97X\cite{wB97X-2008}) were benchmarked against SDSPT2 and experimental results, and the PBE0 functional was chosen based on its satisfactory and uniform accuracy (see Section~\ref{absorption} for details). The orbital diagrams were drawn and visualized with VMD v.1.9.4\cite{VMD}, using cube files generated with the help of Multiwfn v.3.8(dev)\cite{Multiwfn}.
\\\indent
The calculations of ISC rate constants were conducted by the ESD module of the ORCA program, version 5.0.4\cite{orca-5.0,neese2012orca2,neese2018software3,neese2020orca4}, using the thermal vibration correlation function (TVCF) method based on a multimode harmonic oscillator model. Other rate constants involved in the excited state relaxation process were calculated by the MOMAP package, version 2022A\cite{shuai2007,Shuai2008,momap}, again using the TVCF method and a harmonic approximation of the potential energy surfaces. The default parameters of the two programs were used in all TVCF calculations, except for the ``tmax'' parameter in the MOMAP calculations (which controls the propagation time of the TVCF), which was set to 3000 fs. All necessary transition matrix elements, including the transition dipole moments, non-adiabatic coupling matrix elements (NACMEs)\cite{nacme1-2014,nacme2-2014,nacme3-2021}, spin-orbit coupling matrix elements (SOCMEs)\cite{soc-tddft1,nacme2-2014,soc-tddft3}, as well as the seminumerical Hessians necessary for the TVCF calculations, were calculated by BDF. Note however that all NACMEs were computed by U-TDDFT instead of X-TDDFT, since the theory of X-TDDFT NACMEs has not been developed yet; similarly, geometry optimization and frequency calculations of the $^{4}$T$_1$ state were performed at the unrestricted Kohn-Sham (UKS) level, which is justified by the small spin contamination ($\langle S^2 \rangle$ deviation $<$ 0.1) of this state. The ALDA0 noncollinear exchange-correlation (XC) kernel\cite{SF-TDDFT} was used in all spin flip-up Tamm-Dancoff approximation (TDA) calculations (i.e.~calculation of quartet states from a doublet reference), which has proven essential for obtaining correct spin state splittings\cite{XTDDFTbenchmarkDQ}. Duschinsky rotation was considered whenever applicable. The Herzberg-Teller effect was only considered while calculating the radiative relaxation rates, but not the ISC rates, due to program limitations; however this should not change the qualitative conclusions of this paper, since all ISC processes whose Franck-Condon contributions are negligible or zero are expected to contribute negligibly to the photophysics of CuP. Although we have implemented the interface for calculating the Herzberg-Teller effect of phophorescence by BDF and MOMAP, the computation of the geometric derivatives of the doublet-quartet transition dipole moments by finite differences proved to be numerically ill-behaved, as the $M_S=\pm 1/2$ and $M_S=\pm 3/2$ microstates of the $^{4}$T state mix strongly when the geometry is perturbed; note that this phenomenon seems to be related to the involvement of quartet states, since we have never observed similar behavior in triplet phosphorescence rate calculations. We thus estimated the total phosphorescence rate by assuming that the ratios of the Franck-Condon and Herzberg-Teller rates are the same for fluorescence and phosphorescence. This treatment is justified by the observation that the geometries and vibrational frequencies of the $^{2}$T$_1$ and $^{4}$T$_1$ states are very similar.
\\\indent
The active space of the SDSPT2 calculations was selected through the iCAS (imposed automatic selection and localization of complete active spaces) method\cite{iCAS}, and the orbitals were optimized using the iCISCF (iterative configuration interaction (iCI)-based multiconfigurational self-consistent field (SCF) theory) method\cite{iCISCF}, which provided a reference wavefunction for the SDSPT2 calculation. An active space of CAS(13,14) was used in this study. The B-band, Q-band and CT states involved in the excited state relaxation process mainly involve the Cu 3d and 4d orbitals, plus the four porphyrin $\pi$ orbitals of the Gouterman four-orbital model\cite{Gouterman1959study}, making a minimal active space of CAS(13,14). The chosen active space thus properly describes the primary excited states of interest for investigation. Expanding the active space further would result in unnecessary computational overhead without providing additional insights. All SDSPT2 calculations reported herein include the Pople correction.

\section{Results and discussion}

\subsection{Absorption process} \label{absorption}

As is well-known, density functionals generally have difficulties with simultaneously describing local excitation (LE) and CT states with good accuracy. Since we could only afford to do the geometry optimizations and frequency calculations under the DFT and TDDFT levels, a suitable functional that qualitatively reproduces the SDSPT2 excitation energies has to be chosen by comparing the TDDFT vertical absorption energies of a few common functionals with SDSPT2 data. B3LYP and PBE0 are generally common choices for the excited states of metalloporphyrins, and BP86 is often used to optimize their ground-state structures. Pure functionals usually tend to underestimate excitation energies, but empirically, their description of the Q band (an LE state) is better than hybrid functionals, as will be confirmed by our calculation results. As CT states are involved in the relaxation process of the excited states of copper porphyrin, range-separated hybrid functionals (which provide good descriptions of CT states in general) may prove to be suitable as well. These considerations gave a list of four representative functionals, BP86, B3LYP, PBE0 and $\omega$B97X, that were subjected to benchmark calculations.
\\\indent
Different functionals display distinct behaviors for the excitation energies of CuP compared to the results obtained from SDSPT2, as shown in Figure \ref{TDDFT_SDSPT2_comparison}. The two characteristic absorption bands of the porphyrin molecule correspond to the $^{2}$S$_1$ (Q band) and $^{2}$S$_2$ (B band) states, which are the only bright states of most porphyrin complexes in the visible region. They are also the only excited states for which accurate experimental vertical absorption energies are available: in benzene they have been measured as 2.25 and 3.15 eV, respectively\cite{eastwood1969porphyrins}. Moreover, the absorption energy of the $^{2}$T$_1$ state has been measured by fluorescence excitation spectra experiments, but only for certain substituted porphyrins: for example, the $^{2}$T$_1$ absorption energy of CuEtio (Etio = etioporphyrin I) was measured in $n$-octane as 1.81 eV, while the emission energy from the same state in the same solvent was 1.79 eV\cite{eastwood1969porphyrins}. Assuming that the Stokes shift of the $^{2}$T$_1$ state is independent of the porphyrin substituents, and combined with the experimental emission energy of the $^{2}$T$_1$ state of CuP in the same solvent (1.88 eV)\cite{eastwood1969porphyrins}, we obtain an estimate of the experimental $^{2}$T$_1$ absorption energy of CuP as 1.90 eV. Gratifyingly, the SDSPT2 excitation energies of all three states agree with the experimental values to within 0.2 eV, which is typical of the accuracy of SDSPT2\cite{Song_SDS} and confirms the suitability of SDSPT2 as a benchmark reference for CuP. The BP86 functional performs better for these two states, with results closer to the SDSPT2 calculations, suggesting its suitability for localized excitations in the porphyrin system. However, the BP86 functional performs poorly in describing the dark charge transfer (CT) states, significantly underestimating their energies, as expected. In contrast, the range-separated functional $\omega$B97X shows good agreement with the CT states compared to SDSPT2 results, accurately reproducing their energies. However, the $\omega$B97X functional's description of the LE states ($^{2}$S$_1$ and $^{2}$S$_2$) is rather poor, with energies notably higher than the SDSPT2 results. The PBE0 and B3LYP functionals represent compromises between the two kinds of functionals and provide more accurate overall descriptions of the LE and CT states, giving results closer to the SDSPT2 calculations. Considering the overall performance in describing different states, the PBE0 functional slightly outperforms B3LYP, leading to its selection for the remaining part of the present study.
\\\indent
The $^{2}$S$_1$ and $^{2}$S$_2$ states are almost spin-adapted states with minimal spin contamination, even at the U-TDDFT level (Table \ref{GS_TDDFT}), since they are dominated by singdoublet excitations.
As shown in Figure \ref{TDDFT_SDSPT2_comparison}, both X-TDDFT and U-TDDFT provide similar descriptions for these two states; note however that functionals with large amounts of HF exchange generally overestimate the excitation energies of these two states, especially $^{2}$S$_2$. At the TDDFT levels, the CT states are dominated by CO-type excitations (from $\pi$ to 3d$_{x^2-y^2}$), which are also spin-adapted. Both U-TDDFT and X-TDDFT show comparable performance in describing the CT states. However, both methods display large errors compared to SDSPT2 for the CT states. Table \ref{GS_TDDFT} presents the excitation energies and the corresponding dominant excited state compositions, computed at the ground state structure of CuP. It can be observed that the CT states are predominantly composed of double excitations, which are not accurately captured by single-reference methods. Despite this, functionals with large amounts of HF exchange still perform notably better, as is generally expected for CT states. The $^{2}$T$_1$ and $^{2}$T$_2$ states correspond to tripdoublet excitations (from $\pi$ to $\pi$$^*$), and they suffer from significant spin contamination at the U-TDDFT level, since instead of pure doublets, U-TDDFT can only describe these tripdoublet states as a heavy mixture of doublets and quartets, e.g.:
\begin{equation} \label{2T1_determinant}
\Psi(^2\rm{T}_1)^{\rm{U-TDDFT}} \approx -\sqrt{\frac{1}{3}}\Psi(^2\rm{T}_1)^{\rm{X-TDDFT}} + \sqrt{\frac{2}{3}}\Psi(^4\rm{T}_1, \mathit{M_S}=1/2)^{\rm{X-TDDFT}},
\end{equation}
as follows from Eqs.~\ref{Psi_mixed}-\ref{Psi_quartet}. U-TDDFT thus systematically underestimates the excitation energies of the $^{2}$T$_1$ and $^{2}$T$_2$ states, since the energies of quartets are in general lower than the corresponding tripdoublets, as discussed in the Introduction. In Section~\ref{sec:rates} we will also see that part of the underestimation is due to the failure of U-TDDFT to reproduce the energy degeneracy of $\Psi(^4\rm{T}_1, \mathit{M_S}=1/2)$ and $\Psi(^4\rm{T}_1, \mathit{M_S}=3/2)$. On the other hand, X-TDDFT avoids spin contamination through implicitly incorporating extra double excitations necessary for spin-adapting the tripdoublet states (Eq.~\ref{Psi_tripdoublet}), and therefore performs systematically better than U-TDDFT for all the functionals studied herein. The improvements of the excitation energies ($\sim$ 0.05 eV) may seem small, but have profound influences on the magnitude and even the sign of the $^{2}$T$_1$-$^{4}$T$_1$ gap, and therefore on the ratio of fluorescence and phosphorescence emission, as will be detailed in Section~\ref{sec:rates}.
\\\indent
Already from the calculated absorption energies, one can draw some conclusions about the photophysical processes of CuP. The vertical absorption energies of the $^{2}$T$_1$, $^{2}$S$_1$, $^{2}$CT$_1$, $^{2}$CT$_2$ an $^{2}$S$_2$ states of CuP have the intriguing property of being roughly equidistant with very small spacings (0.2-0.4 eV), and the $^{2}$T$_2$ state is furthermore nearly degenerate with $^{2}$S$_1$. Therefore, once CuP is excited to the bright $^{2}$S$_1$ or $^{2}$S$_2$ states by visible light, the molecule is expected to undergo a cascade of ultrafast IC processes, all the way till the lowest doublet state, $^{2}$T$_1$. The availability of an ultrafast IC cascade also means the ISC from these high-lying excited states are probably unimportant, especially considering that copper is a relatively light element. These findings are in qualitative agreement with the experimental observation that the $^{2}$S$_2$ states of substituted copper(II) porphyrins relax to the $^{2}$T$_1$ states in gas phase through a two-step process via the intermediacy of a CT state, with time constants 65 fs and 350-2000 fs, respectively, depending on the substituents\cite{ha2013efficient}. In solution, the $^{2}$S$_1$ state of Cu(II) protoporphyrin IX dimethyl ester was known to relax to $^{2}$T$_1$ within 8 ps\cite{Kobayashi_CuP_AgP}, and for CuTPP as well as CuOEP the same relaxation was also found to occur within the picosecond timescale\cite{Gouterman_SnPdCu}. Recently, the decay rates of the $^{2}$S$_1$ state were measured as 50 fs and 80 fs for CuTPP and CuOEP, respectively, in cyclohexane\cite{Chergui_CuP}. The $^{2}$S$_1$ state lifetime of CuP itself was also estimated, although indirectly from the natural width of the 0-0 peak of the Q band, as 30 fs\cite{Noort_CuP_1976}. Quantitative computation of these IC rates is however beyond the scope of the paper, as the narrow energy gaps and possible involvement of conical intersections probably necessitate nonadiabatic molecular dynamics simulations. Nevertheless, a $^{2}$S$_2\to^{2}$CT$\to^{2}$S$_1/^{2}$T$_2\to^{2}$T$_1$ IC pathway can still be tentatively proposed based on the energy ordering alone. Finally, it is worth noting that the use of the accurate SDSPT2 method, as opposed to TDDFT, is crucial for obtaining a reliable estimate of the qualitative trend of the excited state energies. BP86 predicts that the CT states lie below the $^{2}$T states, leading to a qualitatively wrong IC pathway; $\omega$B97X, on the other hand, grossly overestimates the energy of $^{2}$S$_2$ and would underestimate its tendency to undergo IC to the CT states (Figure \ref{TDDFT_SDSPT2_comparison}). While B3LYP and PBE0 predict reasonable excited state orderings, their accuracy for the CT states cannot be expected in advance without the input of a higher-level computational method, due to the lack of experimental data of the CT states as well as the presence of double excitation contributions in the CT states, which cannot be correctly described under the adiabatic TDDFT framework.
\\
\begin{figure}
	\centering
	\includegraphics[width=0.8\textwidth]{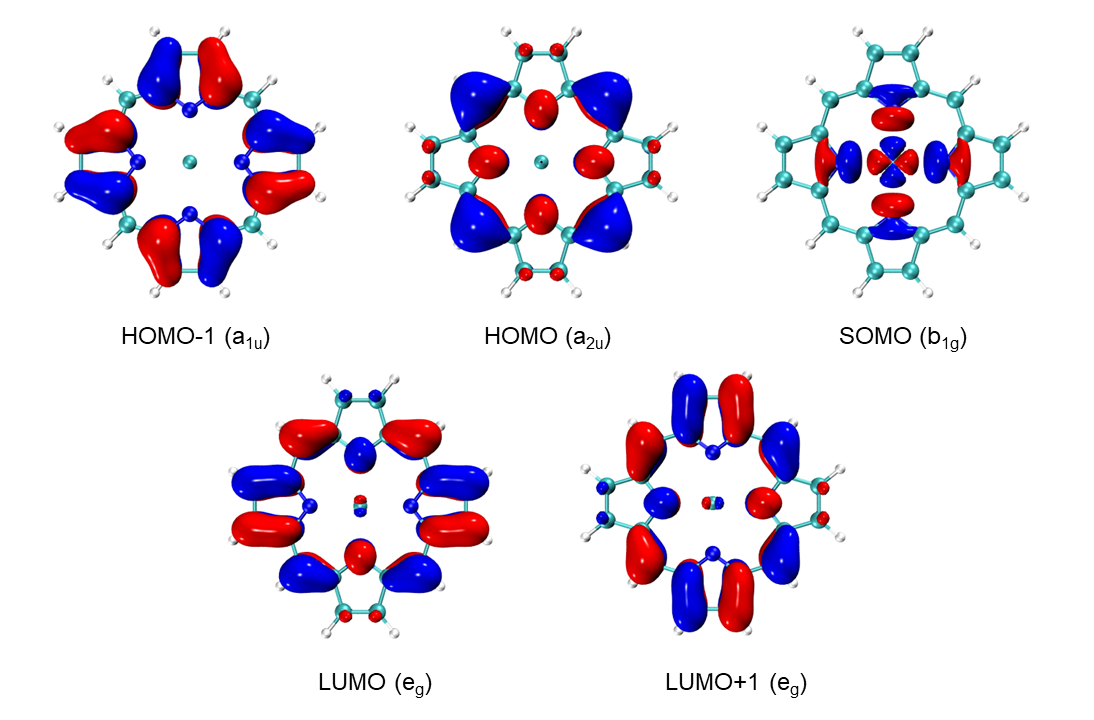}
	\caption{ROKS frontier molecular orbitals of CuP, computed at the sf-X2C-PBE0/x2c-SVPall level of theory.}
	\label{CuP_orbitals}
\end{figure}
\\
\renewcommand{\arraystretch}{1.5}
\begin{table} 
	\centering
	\captionsetup{skip=6pt}
	\caption{The SDSPT2 excitation energies (in eV) computed at the sf-X2C-PBE0/x2c-SVPall ground state structure of CuP, along with the corresponding excited state compositions. $\Delta$$\langle$S$^2$$\rangle$: difference of the excited state's $\langle$S$^2$$\rangle$ value with the ground state $\langle$S$^2$$\rangle$, computed at the U-TD-PBE0 level. Transitions in square brackets represent double excitations.
		}
	\begin{tabular}{cccc}
		\hline
		State & $\Delta$E & $\Delta$$\langle$S$^2$$\rangle$ & Dominant transitions \\
		\hline
		$^{2}$T$_1$ & 2.08 & 1.9994 & $\pi$(a$_{2u}$) $\rightarrow$ $\pi$$^*$(e$_{g}$)  87.1$\%$                                                        \\ 
		$^{2}$T$_2$ & 2.30 & 1.9968 &$\pi$(a$_{1u}$) $\rightarrow$ $\pi$$^*$(e$_{g}$)  86.7$\%$                                                        \\
		$^{2}$S$_1$ & 2.37 & 0.0031 &$\pi$(a$_{1u}$) $\rightarrow$ $\pi$$^*$(e$_{g}$)  56.9$\%$, $\pi$(a$_{2u}$) $\rightarrow$ $\pi$$^*$(e$_{g}$)  36.1$\%$    \\
		$^{2}$CT$_1$ & 2.73 & 0.0101 & [$\pi$(a$_{2u}$) $\rightarrow$ Cu 3d$_{x^2-y^2}$(b$_{1g}$) + Cu 3d$_{xz}$/3d$_{yz}$(e$_{g}$) $\rightarrow$ $\pi$$^*$(e$_{g}$)] 51.0$\%$                                                      \\
		 & & &  $\pi$(a$_{2u}$) $\rightarrow$ Cu 3d$_{x^2-y^2}$(b$_{1g}$)  39.6$\%$                                                    \\
		$^{2}$CT$_2$ & 2.93 & 0.0064 &[$\pi$(a$_{1u}$) $\rightarrow$ Cu 3d$_{x^2-y^2}$(b$_{1g}$) + Cu 3d$_{xz}$/3d$_{yz}$(e$_{g}$) $\rightarrow$ $\pi$$^*$(e$_{g}$)] 42.4$\%$                                                 \\
		 & & & $\pi$(a$_{1u}$) $\rightarrow$ Cu 3d$_{x^2-y^2}$(b$_{1g}$)  34.5$\%$        
		                                                              \\ 
		$^{2}$S$_2$ & 3.30 & 0.0115 &$\pi$(a$_{2u}$) $\rightarrow$ $\pi$$^*$(e$_{g}$)  52.5$\%$ , $\pi$(a$_{1u}$) $\rightarrow$ $\pi$$^*$(e$_{g}$)  31.8$\%$    \\ 
		\hline
	\end{tabular}
\label{GS_TDDFT}
\end{table}		
\begin{figure}[htbp]
	\centering
	\includegraphics[scale=0.2]{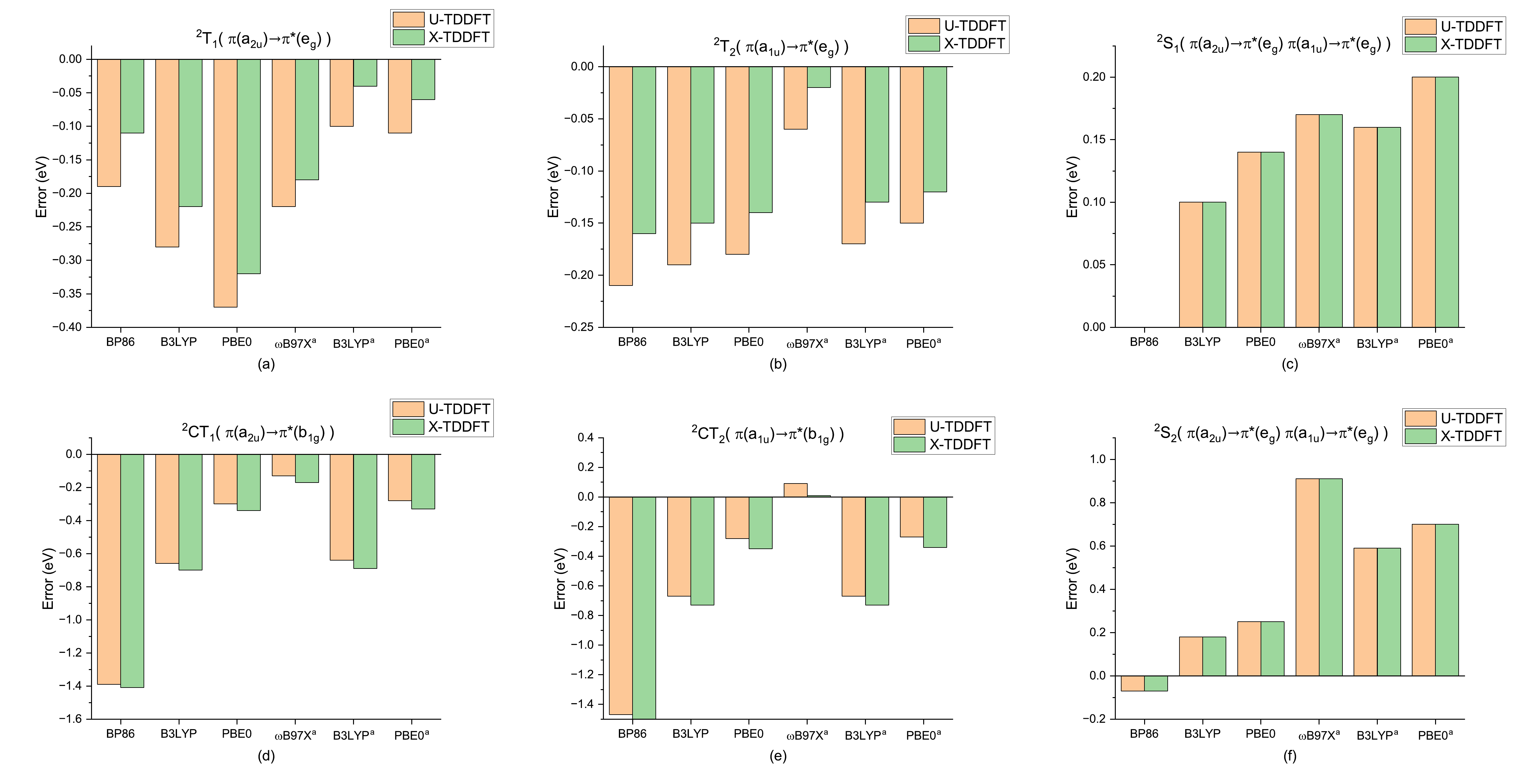}
	\caption{Errors of different excited states of CuP with respect to SDSPT2 values. $^{a}$TDA}
	\label{TDDFT_SDSPT2_comparison}
\end{figure}
\\
\subsection{Analysis of the equilibrium geometries of the first doublet excited state}
Since all higher lying excited states are predicted to convert to $^{2}$T$_1$ over a short timescale, to study the luminescence of CuP (and probably also other Cu(II) porphyrin complexes bearing alkyl or aryl substituents, given that these substituents do not change excitation energies drastically\cite{eastwood1969porphyrins}), it should suffice to study the radiative and non-radiative processes starting from the $^{2}$T$_1$ state. As $^{2}$T$_1$ is the lowest doublet state, we expect that its lifetime is long enough for it to relax to its equilibrium structure, before any further transitions occur. Therefore, accurately predicting the equilibrium geometry of the $^{2}$T$_1$ state is crucial for subsequent studies.
\\\indent
Some selected bond lengths for the optimized ground state and excited state structures are provided in Table~\ref{equilibrium_gs_es}. The difference in ground state bond lengths between the UKS and ROKS methods is extremely small ($<$ 0.0001 \AA), as can be seen from their root mean square deviation (RMSD), which can be attributed to the extremely small UKS spin contamination of the ground state of CuP ($\langle S^2 \rangle_{\rm{PBE0}} = 0.7532$). The doubly degenerate $^{2}$T$_1$ state, which belongs to the doubly degenerate E$_u$ irreducible representation (irrep) under the D$_{4h}$ group, undergoes Jahn-Teller distortion to give a D$_{2h}$ structure, where two of the opposing Cu-N bonds are elongated but the corresponding pyrrole rings remain almost intact, while the other two Cu-N bonds are almost unchanged but the corresponding pyrrole rings exhibit noticeable deformation. The U-TDDFT and X-TDDFT bond lengths of the $^{2}$T$_1$ state show larger deviations than the UKS and ROKS ground state ones, with the largest deviation exceeding 0.001 \AA{} (the C-C (mn) bond), which is also reflected in the RMSD values. However, the structure differences are still small on an absolute scale. This suggests that the coupling of the unpaired Cu(II) d electron and the porphyrin triplet is weak, so that a reasonable tripdoublet state geometry is obtained even if this coupling is described qualitatively incorrectly (as in U-TDDFT). By contrast, our previous benchmark studies on small molecules (where the coupling between unpaired electrons is much larger) revealed that X-TDDFT improves the U-TDDFT bond lengths by 0.01-0.05 \AA{} on average, depending on the functional and the molecule\cite{XTDDFT,XTDDFTgrad}.

\renewcommand{\arraystretch}{1.5}
\begin{table} 
	\centering
	\captionsetup{skip=6pt}
	\caption{The equilibrium bond lengths (in $\rm{\AA}$) of the ground state ($^2$S$_0$) and the first doublet excited state ($^{2}$T$_1$) of the CuP molecule.}
	\begin{tabular}{cccccc}
		\hline
		State & Cu-N & C-N & C-C (mn)$^a$ & C-C (mt)$^a$ &C-C (st)$^a$ \\
		\hline
		\multicolumn{6}{c}{UKS} \\
		$^2$S$_0$   & 2.0148 & 1.3652 & 1.3881 & 1.4410 & 1.3611 \\	
		\multicolumn{6}{c}{U-TDDFT} \\
		\multirow{2}{*}{$^2$T$_1$} & 2.0190 & 1.3683 & 1.4155 & 1.4145 & 1.3885 \\ 
		& 2.0416 & 1.3674 & 1.3842 & 1.4447 & 1.3584 \\
		\multicolumn{6}{c}{ROKS} \\
		$^2$S$_0$ & 2.0148 & 1.3652 & 1.3881 & 1.4410 & 1.3612 \\
		\multicolumn{6}{c}{X-TDDFT} \\
		\multirow{2}{*}{$^2$T$_1$} & 2.0198 & 1.3681 & 1.4152 & 1.4149 & 1.3886 \\
		& 2.0413 & 1.3668 & 1.3853 & 1.4442 & 1.3590 \\
		RMSD$^b$ &\multicolumn{5}{c}{0.00002} \\ 
		RMSD$^c$ &\multicolumn{5}{c}{0.00102} \\
		\hline
	\end{tabular}
	\\
	\raggedright
		$^a$See Figure~\ref{structures} for the labeling of atoms. \\
		$^b$The RMSD ($\rm{\AA}$) between the optimized $^2$S$_0$ state structures obtained using UKS and ROKS.\\
		$^c$The RMSD ($\rm{\AA}$) between the optimized $^2$T$_1$ state structures obtained using U-TDDFT and X-TDDFT.\\
		\label{equilibrium_gs_es}
\end{table}

\subsection{Relaxation Processes of the $^2$T$_1$ state} \label{sec:rates}
\begin{figure}
	\centering
	\includegraphics[width=0.8\textwidth]{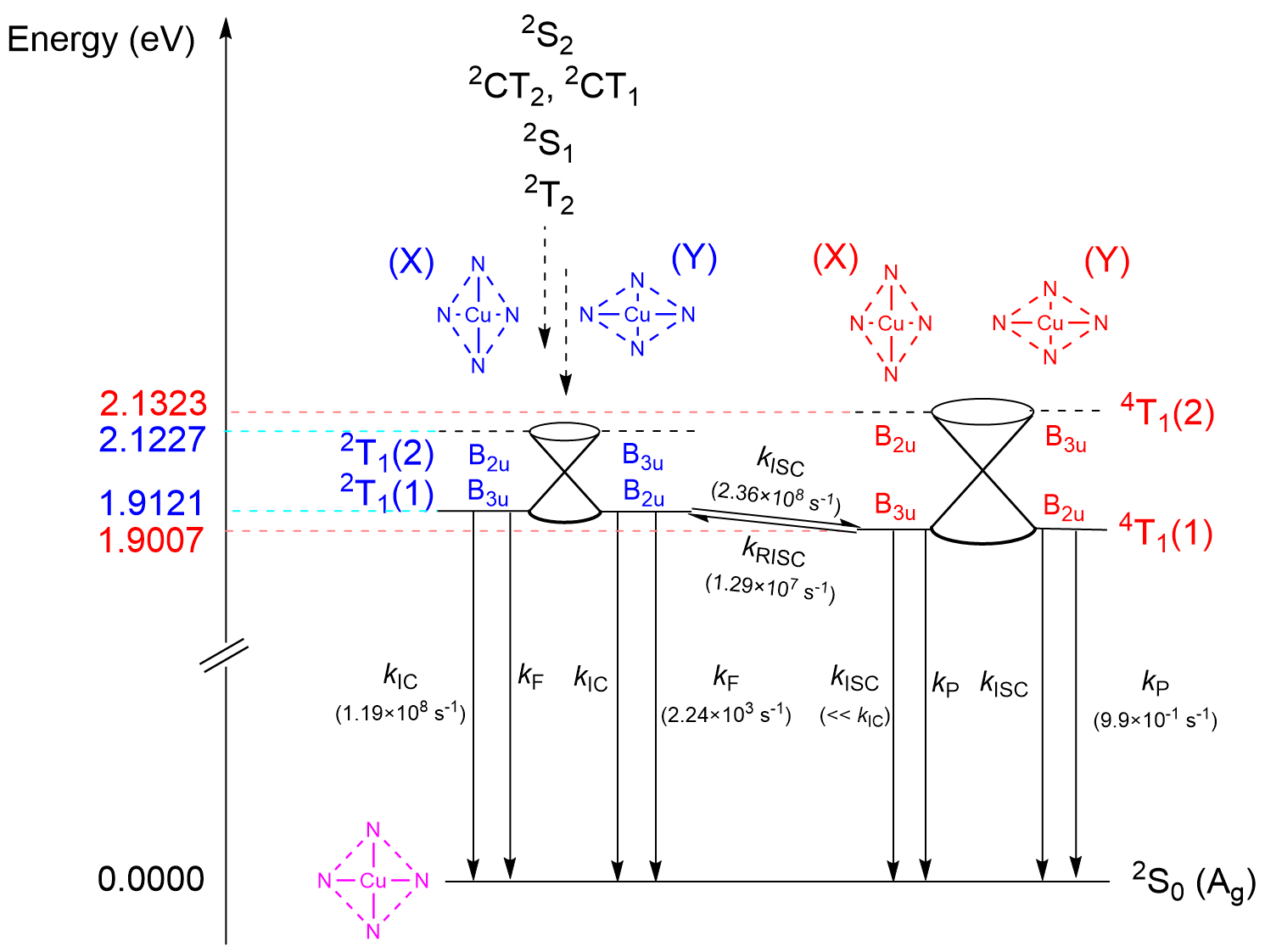}
	\caption{Radiative and non-radiative relaxation pathways of the $^2$T$_1$ state. Both the $^2$T$_1$ and $^4$T$_1$ states are splitted by the Jahn-Teller effect to give two adiabatic states, labeled (1) and (2). Each of the (1) states have two equivalent D$_{2h}$ equilibrium structures, labeled (X) and (Y). The (2) states do not have equilibrium structures and are connected with the corresponding (1) states via conical intersections. The adiabatic excitation energies of the (1) states, as well as the energies of the (2) states at the equilibrium geometries of their corresponding (1) states, are shown on the left. The transition rates are calculated at 83 K in the gas phase. The forward and reverse ISC rates between $^2$T$_1(1)(\rm{X})$ and $^4$T$_1(1)(\rm{Y})$ are equal to those between $^2$T$_1(1)(\rm{Y})$ and $^4$T$_1(1)(\rm{X})$ by symmetry, but the former ISC processes are omitted for clarity. Transition rates that are obviously equal by symmetry reasons are shown only once.}
	\label{Jablonski}
\end{figure}

As revealed by the above analyses, the relaxation process from high-lying excited states to the $^2$T$_1$ state is rapid, and the only energetically accessible relaxation pathways are the radiative (fluorescence) and non-radiative (IC) relaxations from $^2$T$_1$ to the ground state $^2$S$_0$, as well as the ISC from $^2$T$_1$ to $^4$T$_1$. The $^4$T$_1$ state can furthermore convert back to the $^2$T$_1$ state through reverse ISC (RISC), or relax to the ground state via radiative (phosphorescence) or non-radiative (ISC) pathways (Figure~\ref{Jablonski}).
\\\indent
Before we discuss the quantitative values of transition rates, we first analyze the relevant electronic states from the viewpoint of point group symmetry. The equilibrium structures of the $^2$T$_1$ and $^4$T$_1$ states are both distorted owing to the Jahn-Teller effect, and possess only D$_{2h}$ symmetry, compared to the D$_{4h}$ symmetry of the ground state equilibrium structure. The implications are two-fold: the double degeneracy of the $^n$T$_1 (n=2 \rm{~or~} 4)$ state at the D$_{4h}$ geometry (where they both belong to the E$_u$ irrep) is lifted to give two adiabatic states, hereafter termed the $^n$T$_1(1)$ and $^n$T$_1(2)$ states, respectively, where $^n$T$_1(1)$ is the state with the lower energy; and the potential energy surface of the $^n$T$_1(1)$ state has two chemically equivalent D$_{2h}$ minima, $^n$T$_1(1)(\rm{X})$ and $^n$T$_1(1)(\rm{Y})$, where different pairs of Cu-N bonds are lengthened and shortened (see the schematic depictions in Figure~\ref{Jablonski}). Although $^n$T$_1(1)(\rm{X})$ and $^n$T$_1(1)(\rm{Y})$ are on the same adiabatic potential energy surface, their electronic wavefunctions represent different diabatic states, as they belong to the B$_{3u}$ and B$_{2u}$ irreps, respectively. The $^n$T$_1(1)(\rm{X})$ structure is diabatically connected to $^n$T$_1(2)(\rm{Y})$ (i.e.~the $^n$T$_1(2)$ state at the equilibrium structure of $^n$T$_1(1)(\rm{Y})$) via a D$_{4h}$ conical intersection, while $^n$T$_1(1)(\rm{Y})$ is diabatically connected to $^n$T$_1(2)(\rm{X})$ via the same conical intersection. Thus, the $^n$T$_1(2)(\rm{X})$ and $^n$T$_1(2)(\rm{Y})$ states are expected to undergo ultrafast IC from the D$_{4h}$ conical intersection, to give the $^n$T$_1(1)(\rm{Y})$ and $^n$T$_1(1)(\rm{X})$ states as the main products, respectively. The direct transition from $^n$T$_1(2)$ to states other than $^n$T$_1(1)$ can therefore be neglected.
\\\indent
From the irreps of the electronic states, we conclude that certain ISC transitions are forbidden by spatial symmetry. These include the transitions between $^2$T$_1(1)(\rm{X})$ and $^4$T$_1(1)(\rm{X})$, between $^2$T$_1(1)(\rm{Y})$ and $^4$T$_1(1)(\rm{Y})$, and between any one of the $^4$T$_1(1)$ structures and $^2$S$_0$. All IC and radiative transitions, plus the ISC transitions between $^2$T$_1(1)(\rm{X})$ and $^4$T$_1(1)(\rm{Y})$ as well as between $^2$T$_1(1)(\rm{Y})$ and $^4$T$_1(1)(\rm{X})$, are symmetry allowed. While symmetry forbidden ISC processes can still gain non-zero rates from the Herzberg-Teller effect, we deem that the rates are not large enough to have any noticeable consequences. On one hand, the two symmetry forbidden ISC pathways between the $^2$T$_1(1)$ and $^4$T$_1(1)$ states are overshadowed by the two symmetry allowed ones, so that the total ISC rate between $^2$T$_1(1)$ and $^4$T$_1(1)$ is undoubtedly determined by the latter alone. The ISC from $^4$T$_1(1)$ to $^2$S$_0$, on the other hand, has to compete with the IC process from $^2$T$_1(1)$ to $^2$S$_0$ in order to affect the quantum yield or the dominant relaxation pathway of the system noticeably, but the latter process is both spin-allowed and spatial symmetry-allowed, while the former is forbidden in both aspects. We therefore neglect all ISC rates whose Franck-Condon contributions are zero by spatial symmetry.
\\\indent
We then calculated the rate constants for all transitions between $^2$T$_1$, $^4$T$_1$ and $^2$S$_0$ whose rates are non-negligible, by the TVCF method. The rates (Figure~\ref{Jablonski}) were calculated at 83 K, the temperature used in the quantum yield studies of Ref.~\cite{eastwood1969porphyrins}; the latter studies gave a luminescence quantum yield of 0.09, in a solvent mixture of diethyl ether, isopentane, dimethylformamide and ethanol. The accurate treatment of solvation effects is however complicated and beyond the scope of the paper, so that all transition rates were computed in the gas phase.
Our calculated $k_{\rm{ISC}}$ from $^2$T$_1$ to $^4$T$_1$ is only slightly larger than the $k_{\rm{IC}}$ from $^2$T$_1$ to $^2$S$_0$, suggesting that treating the $^2$T$_1$ and $^4$T$_1$ states as a rapid equilibrium (as in, e.g.~Ref.~\cite{Gouterman-Cu-VO-Co-porphyrins-1969} and~\cite{CuP_hightemp_lumin}) is not justified at least in the gas phase. At 83 K, the RISC from $^4$T$_1$ to $^2$T$_1$ is 11 \% of the forward ISC rate. Both rates are in favorable agreement with the experimental values of Cu(II) protoporphyrin IX dimethyl ester in benzene, $k_{\rm{ISC}} = 1.6 \times 10^9 \rm{s}^{-1}$ and $k_{\rm{RISC}} = 5.6 \times 10^8 \rm{s}^{-1}$, at room temperature\cite{Kobayashi_CuP_AgP}.
Our computed ISC and RISC rates give a $^2$T$_1$-to-$^4$T$_1$ equilibrium concentration ratio of 1:9.0 when $k_{\rm{IC}}$ is neglected, but our kinetic simulation shows that the steady state concentration ratio is 1:14.9 when the latter is considered, further illustrating that treating the $^2$T$_1$-$^4$T$_1$ interconversion as a fast equilibrium can lead to noticeable error. Nevertheless, the fluorescence rate of $^2$T$_1$ still exceeds the phosphorescence rate of $^4$T$_1$ by three orders of magnitude, which more than compensates for the low steady state concentration of $^2$T$_1$. Similar conclusions could be derived from the rates reported in Ref.~\cite{Gouterman-Cu-VO-Co-porphyrins-1969} ($3.6\times 10^3 \rm{s}^{-1}$ and $8.3\times 10^{-1} \rm{s}^{-1}$, respectively), calculated from semiempirical exchange and SOC integrals and experimental absorption oscillator strengths, which agree surprisingly well with the rates that we obtained here.
Kinetic simulation suggests that 99.6 \% of the total luminescence at this temperature is contributed by fluorescence, and only 0.4 \% is due to phosphorescence. This can be compared with the experimental finding by Bohandy and Kim\cite{CuP_hightemp_lumin} that the phosphorescence of CuP at 86 K is observable as a minor 0-0 peak besides the 0-0 fluorescence peak, with a fluorescence to phosphorescence ratio of about 5:1 to 10:1 (as estimated from Fig.~5 of Ref.~\cite{CuP_hightemp_lumin}); however note that this study was performed in a triphenylene solid matrix.
\\\indent
The total luminescence quantum yield is predicted by our kinetic simulations to be $1.9\times 10^{-5}$, three orders of magnitude smaller than the experimental quantum yield (0.09) in solution. We believe one possible reason is that the $^2$T$_1$-$^4$T$_1$ gap of CuP is larger in solution than in the gas phase. This can already be seen from the experimental $^2$T$_1$-$^4$T$_1$ 0-0 gaps of CuP in solid matrices with different polarities: the 0-0 gap was measured in polymethylmethacrylate as 500 cm$^{-1}$\cite{Gouterman_CuP_phosphorescence_1968}, but 310-320 cm$^{-1}$ in $n$-octane\cite{Noort_CuP_1976} and 267 cm$^{-1}$ in triphenylene\cite{CuP_hightemp_lumin}. Therefore, the 0-0 gap in the gas phase is probably smaller than 267 cm$^{-1}$, and indeed, our X-TDDFT calculations predict an adiabatic $^2$T$_1$-$^4$T$_1$ gap of 92 cm$^{-1}$ in the gas phase. The larger $^2$T$_1$-$^4$T$_1$ gap in solution compared to the gas phase is expected to introduce a Boltzmann factor of $\exp\left(-(E_{\rm{sol}}-E_{\rm{gas}})/RT\right)$ to $k_{\rm{RISC}}$, while changing the other rates negligibly. Setting $E_{\rm{sol}}=$267 cm$^{-1}$ and $E_{\rm{gas}}=$92 cm$^{-1}$, we obtain a solution phase $k_{\rm{RISC}}$ of $9.02\times 10^5 \rm{s}^{-1}$, from which kinetic simulations give a fluorescence-phosphorescence ratio of 12:1, in quantitative agreement with experiment\cite{CuP_hightemp_lumin}.
Setting $E_{\rm{sol}}=$500 cm$^{-1}$ (as appropriate for the polar solvent used in Ref.~\cite{eastwood1969porphyrins}) gives $k_{\rm{RISC}}=1.60\times 10^4 \rm{s}^{-1}$, and a total luminescence quantum yield of $1.1\times 10^{-4}$, with 18 \% contribution from fluorescence and 82 \% from phosphorescence. The remaining discrepancy ($\sim$ 800x) of the experimental and calculated quantum yields can be attributed to the restriction of the molecular vibrations of CuP by the low temperature (and thus viscous) solvent, which is expected to suppress the IC process significantly.
\\\indent
Interestingly, U-TDA completely fails to reproduce the qualitative picture of Figure~\ref{Jablonski} and predicts a $^2$T$_1$-$^4$T$_1$ adiabatic gap of the wrong sign (-276 cm$^{-1}$), violating Hund's rule. At first sight, this may seem surprising: since the U-TDA ``tripdoublet state'' is a mixture of the true tripdoublet state and the quartet state, the U-TDA $^2$T$_1$ energy should lie in between the energies of the true $^2$T$_1$ state and the $^4$T$_1$ state, which means that the U-TDA $^2$T$_1$-$^4$T$_1$ gap should be smaller than the X-TDA gap but still have the correct sign. However, the U-TDA $^2$T$_1$ state is contaminated by the $\mathit{M_S}=1/2$ component of the $^4$T$_1$ state (Eq.~\ref{2T1_determinant}), while a spin flip-up U-TDA calculation of the $^4$T$_1$ state gives its $\mathit{M_S}=3/2$ component. The two spin components obviously have the same energy in the exact non-relativistic theory and in all rigorous spin-adapted methods, but not in U-TDA, even when the ground state is not spin-contaminated\cite{SF-TDDFT,XTDDFTbenchmarkDQ}. This shows that the restoration of the degeneracy of spin multiplets by the random phase approximation (RPA) correction in X-TDDFT\cite{XTDDFT} indeed leads to qualitative improvement of the excitation energies, instead of being merely a solution to a conceptual problem. It also shows that estimating the tripdoublet energy by extrapolating from the energies of the spin-contaminated tripdoublet and the quartet by e.g.~the Yamaguchi method\cite{Yamaguchi_spin_projection} does not necessarily give a qualitatively correct estimate of the spin-pure tripdoublet energy. The inverted doublet-quartet gap introduces qualitative defects to the computed photophysics of CuP. Already when the doublet-quartet gap is zero, the Boltzmann factor is expected to raise the $k_{\rm{RISC}}$ to $9.24\times 10^7 \rm{s}^{-1}$, reducing the ratio of phosphorescence in the total luminescence to 0.08 \%. Further raising the quartet to reproduce the U-TDA doublet-quartet gap will reduce the $k_{\rm{ISC}}$ to $1.42\times 10^6 \rm{s}^{-1}$, which reduces the ratio of phosphorescence to 0.0007 \%. These values are obviously in much worse agreement with the experiments\cite{CuP_hightemp_lumin}.
\\\indent
Finally, we briefly comment on the luminescence lifetimes. The luminescence of CuP is known to decay non-exponentially\cite{Gouterman_CuP_phosphorescence_1968}, so its luminescence lifetime can only be approximately determined. The luminescence lifetime of CuP has been determined as 400 $\mu$s\cite{Gouterman_CuP_phosphorescence_1968} at 80 K in polymethylmethacrylate, and a biexponential decay with lifetimes 155 and 750 $\mu$s was reported\cite{eastwood1969porphyrins} at 78 K in methylphthalylethylglycolate. The same references also reported that the luminescence lifetimes of CuOEP and CuTPP are also within the 50-800 $\mu$s range. However, in a room temperature toluene solution the luminescence lifetimes of CuOEP and CuTPP were reported to be 115 and 30 ns, respectively\cite{base_quench_CuP}, and a few nanoseconds in the gas phase\cite{ha2013efficient}. If we define the luminescence lifetime as the time needed for $1-1/e \approx 63.2\%$ of the luminescence to be emitted, then kinetic simulations from our X-TDDFT rate constants give a gas-phase luminescence lifetime of 70 ns at 83 K, which is much shorter than the low-temperature condensed phase results but in very good agreement with the room-temperature solution phase experiments. The fact that the computed lifetime is one order of magnitude longer than the gas-phase experimental result is probably due to the $^2$T$_1$ state being vibrationally excited in the experimental study. However, using the ISC and RISC rate constants consistent with the U-TDA doublet-quartet gap, one obtains a lifetime of 8.5 fs, which seems somewhat too short given that the highly vibrationally excited CuP molecules in Ref.~\cite{ha2013efficient}, which carry all the excess thermal energy ($>$ 1 eV) after the IC process from B band excitation, should have a much shorter lifetime than a ``cold'' CuP molecule at 83 K. Thus, our results suggest that X-TDDFT/X-TDA seems to give a more accurate luminescence lifetime than U-TDDFT/U-TDA, and also confirm that the discrepancy of the experimental and calculated quantum yields is probably due to suppression of the IC of $^2$T$_1$ by the low temperature solvent.
\\
\\
\subsection{Discussions}
\begin{figure}[htbp]
	\centering
	\includegraphics[scale=0.6]{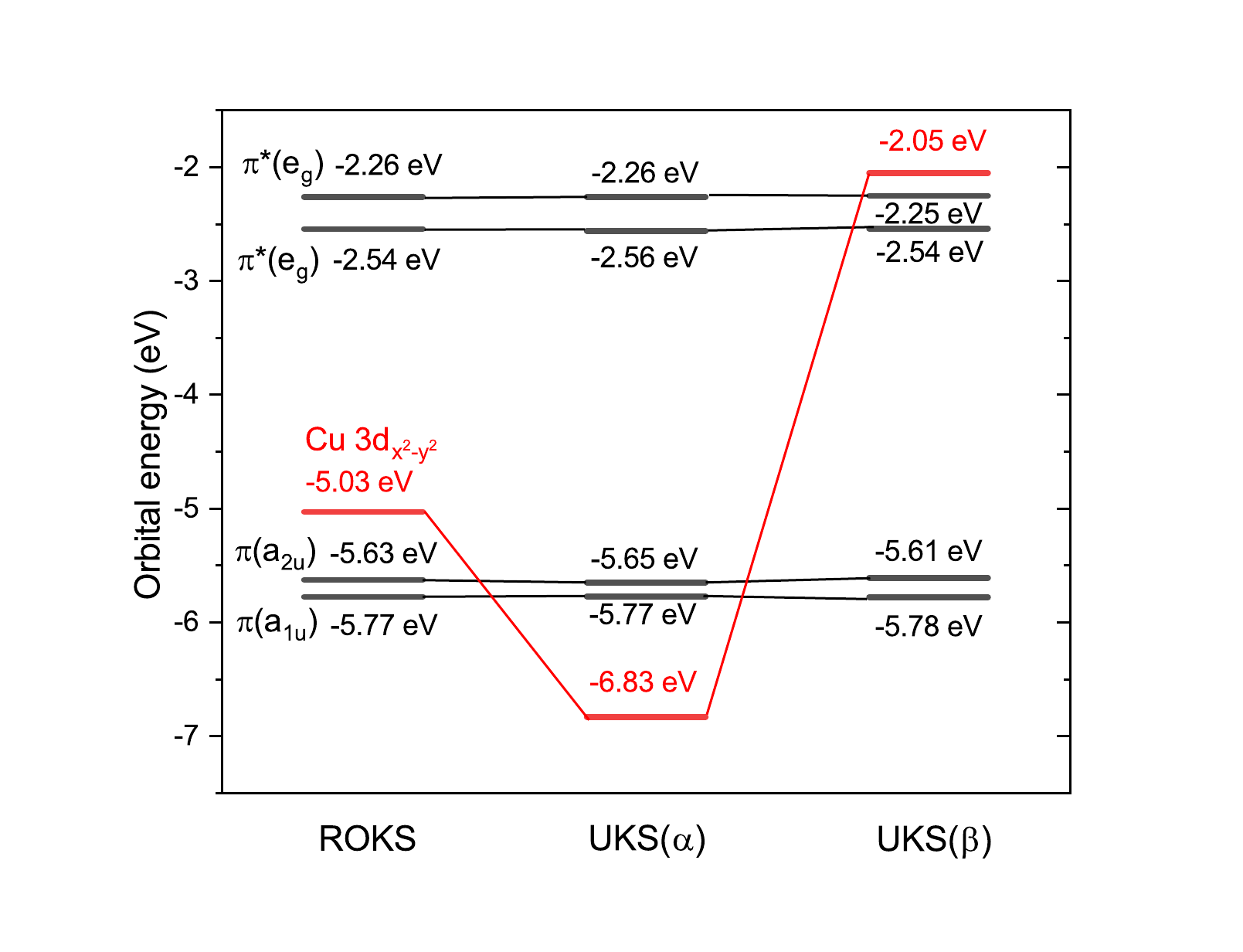}
	\caption{ROKS and UKS orbital energies of CuP at the X-TDDFT and U-TDDFT $^2$T$_1$ equilibrium geometries, respectively, computed at the sf-X2C-PBE0/x2c-SVPall level of theory.}
	\label{orbital_energies}
\end{figure}

As mentioned in the Introduction, the simple orbital energy difference model based on a restricted open-shell determinant (Figure~\ref{CO_OV_CV}) predicts that the lowest tripdoublet is at least the third lowest doublet excited state of any doublet molecule (as long as the ROKS ground state satisfies the \emph{aufbau} rule), since the tripdoublet is higher than at least one CO state and at least one OV state. It therefore comes as a surprise that the lowest three spin-conserving excited states of CuP are all tripdoublets (Table~\ref{GS_TDDFT}), not to mention that all of them are doubly degenerate, even though the ROKS ground state of CuP is indeed an \emph{aufbau} state (Figure~\ref{orbital_energies}). This suggests a failure of the ROKS orbital energy difference model.
\\\indent
To understand why the ROKS orbital energies fail qualitatively for describing the excited state ordering of CuP, despite that the X-TDDFT method (which uses the ROKS determinant as the reference state) still gives reasonable excitation energies as compared to SDSPT2, we note that the $\alpha$ and $\beta$ Fock matrices of an ROKS calculation are in general not diagonal under the canonical molecular orbital (CMO) basis. Only the unified coupling operator $\mathbf{R}$, assembled from blocks of the CMO Fock matrices,
\begin{equation}
\mathbf{R} = \left(\begin{array}{ccc}\frac{1}{2}(\mathbf{F}_{\mathrm{CC}\alpha}+\mathbf{F}_{\mathrm{CC}\beta}) & \mathbf{F}_{\mathrm{CO}\beta} & \frac{1}{2}(\mathbf{F}_{\mathrm{CV}\alpha}+\mathbf{F}_{\mathrm{CV}\beta}) \\ \mathbf{F}_{\mathrm{OC}\beta} & \frac{1}{2}(\mathbf{F}_{\mathrm{OO}\alpha}+\mathbf{F}_{\mathrm{OO}\beta}) & \mathbf{F}_{\mathrm{OV}\alpha} \\ \frac{1}{2}(\mathbf{F}_{\mathrm{VC}\alpha}+\mathbf{F}_{\mathrm{VC}\beta}) & \mathbf{F}_{\mathrm{VO}\alpha} & \frac{1}{2}(\mathbf{F}_{\mathrm{VV}\alpha}+\mathbf{F}_{\mathrm{VV}\beta}) \end{array}\right), \label{ROKS_Fock}
\end{equation}
is diagonal\cite{Nakatsuji-ROHF}. Note that herein we have used the Guest-Saunders parameterization\cite{GuestSaunders} of the diagonal blocks of $\mathbf{R}$, which is the default choice of the BDF program, although our qualitative conclusions are unaffected by choosing other parameterizations. However, the leading term of the X-TDDFT calculation is not simply given by the eigenvalue differences of $\mathbf{R}$,
\begin{equation}
\Delta_{ia\sigma,jb\tau}' = \delta_{\sigma\tau}\delta_{ij}\delta_{ab}(R_{aa} - R_{ii}), \label{Delta_R}
\end{equation}
but rather from the $\alpha$ and $\beta$ Fock matrices themselves via
\begin{equation}
\Delta_{ia\sigma,jb\tau} = \delta_{\sigma\tau}(\delta_{ij}F_{ab\sigma} - \delta_{ab}F_{ji\sigma}). \label{Delta_F}
\end{equation}
Here $i,a$ represent occupied CMOs, $j,b$ virtual CMOs, and $\sigma,\tau$ spin indices. For the diagonal matrix element of an arbitrary single excitation, Eq.~\ref{Delta_F} and Eq.~\ref{Delta_R} differ by the following term:
\begin{equation}
\Delta_{ia\sigma,ia\sigma} - \Delta_{ia\sigma,ia\sigma}' = \frac{1}{2}\left((F_{aa\sigma} - F_{aa\sigma'}) - (F_{ii\sigma} - F_{ii\sigma'})\right), \label{Delta_diff}
\end{equation}
where $\sigma'$ is the opposite spin of $\sigma$. For a general hybrid functional, the Fock matrix element differences in Eq.~\ref{Delta_diff} are given by (where $p$ is an arbitrary CMO, $c_{\rm{x}}$ is the proportion of HF exchange, and $v^{\rm{xc}}$ is the XC potential)
\begin{equation}
F_{pp\beta} - F_{pp\alpha} = c_{\rm{x}}(pt|pt) + (v_{pp\beta}^{\rm{xc}} - v_{pp\alpha}^{\rm{xc}}), \label{F_diff}
\end{equation}
assuming, for the sake of simplicity, that there is only one open-shell orbital $t$ in the reference state. Assuming that the XC potential behaves similarly as the exact exchange potential, the difference Eq.~\ref{F_diff} is positive, and should usually be the largest when $p=t$, while being small when $p$ is spatially far from $t$. The corollary is that the orbital energy difference approximation Eq.~\ref{Delta_R} should agree well with the X-TDDFT leading term Eq.~\ref{Delta_F} for CV excitations (where the difference is proportional to the small exchange integral $(pt|pt)$), but underestimate the excitation energies of CO and OV excitations by a correction proportional to the large $(tt|tt)$ integral.
\\\indent
The underestimation of CO and OV excitation energies by ROKS orbital energy differences opens up the possibility of engineering a system to break the $\omega_{ia} > \max( \omega_{it}, \omega_{ta})$ constraint inherent in the ROKS orbital energy difference model, and make the lowest doublet excited state a tripdoublet. Possible approaches include:
\begin{enumerate}
	\item Increase the difference Eq.~\ref{Delta_diff} for the CO and OV states, while keeping it small for the lowest CV state, so that all CO and OV states are pushed above the lowest CV state. This is most easily done by making the open-shell orbital $t$ very compact, which naturally leads to a larger $F_{tt\beta} - F_{tt\alpha}$ (due to a larger $(tt|tt)$) but a smaller $F_{pp\beta} - F_{pp\alpha}, p\in\{i,a\}$ (due to a small absolute overlap between the $p$ and $t$ orbitals).
	\item Reduce the orbital energy gap between the highest doubly occupied orbital and the lowest unoccupied orbital, which also helps to reduce the excitation energy of the lowest CV state. However, a too small orbital energy gap will favor the IC of the tripdoublet to the ground state, which may quench the fluorescence of the tripdoublet. As already mentioned in Section~\ref{sec:rates}, the IC rate of CuP is already large enough to make CuP only barely fluorescent (quantum yield $\sim 10^{-5}$) in the gas phase, and a viscous solvent seems to be required to suppress the IC contribution and make the fluorescence stronger.
\end{enumerate}
Now, it becomes evident that CuP fits the above design principles very well. The unpaired electron in the ground state of CuP is on the Cu 3d$_{x^2-y^2}$ orbital (Figure~\ref{CuP_orbitals}), which is spatially localized. Moreover, the Cu 3d$_{x^2-y^2}$ orbital occupies a different part of the molecule than the ligand $\pi$ and $\pi^*$ orbitals, which results in a small absolute overlap between the orbitals and helps to reduce the effect of Eq.~\ref{Delta_diff} on the CV excitation energies. To quantitatively assess the effect of Eq.~\ref{Delta_diff} on the CO and OV excitation energies, we note that the X-TDDFT leading term Eq.~\ref{Delta_F} is nothing but the UKS orbital energy difference, if the shape differences of the UKS and ROKS orbitals are neglected. Therefore, we have plotted the UKS orbital energies of CuP in Figure~\ref{orbital_energies} as well. Intriguingly, the $\alpha$ Cu 3d$_{x^2-y^2}$ orbital now lies below the porphyrin $\pi$(a$_{1u}$) and $\pi$(a$_{2u}$) orbitals, while the $\beta$ Cu 3d$_{x^2-y^2}$ orbital lies above the porphyrin $\pi^*$(e$_g$) orbitals. Therefore, the differences of UKS orbital energies predict that the lowest excited states of CuP are the CV states obtained from exciting an electron from $\pi$(a$_{1u}$) and $\pi$(a$_{2u}$) to $\pi^*$(e$_g$). This is not only consistent with our U-TD-PBE0 excitation energies, but also the X-TD-PBE0 and SDSPT2 results (save for the $^2$S$_2$ state at all the three levels of theory, as well as the $^2$S$_1$ state at the TDDFT level, which are higher than the CO-type CT states computed at the respective levels of theory), despite that the latter methods are spin-adapted. Note also that although the $(tt|tt)$ integral leads to a huge splitting between the $\alpha$ and $\beta$ Cu 3d$_{x^2-y^2}$ orbitals, the splitting is only barely enough for the UKS orbital energy differences to predict a tripdoublet first excited state: if the $\beta$ Cu 3d$_{x^2-y^2}$ orbital were just 0.2 eV lower, one would predict that the CO-type CT excitation $\pi$(a$_{2u}$)$\rightarrow$Cu 3d$_{x^2-y^2}$ is lower than the lowest tripdoublet $\pi$(a$_{2u}$)$\rightarrow$$\pi^*$(e$_g$). This can be compared with the 0.65 eV gap between the SDSPT2 $^2$T$_1$ and $^2$CT$_1$ states, computed at the ground state structure of CuP (Table~\ref{GS_TDDFT}). Alternatively, one may say that the HOMO-LUMO gap of the porphyrin ligand is barely narrow enough to fit within the energy window between the $\alpha$ and $\beta$ Cu 3d$_{x^2-y^2}$ orbitals, which clearly illustrates the importance of using a narrow-gap ligand for designing systems with a tripdoublet first excited state.
\\\indent
To conclude this section, we briefly note that making the first doublet excited state a tripdoublet state does not guarantee the realization of tripdoublet fluorescence. Two remaining potential obstacles are (1) the IC of the tripdoublet state to the ground state and (2) the ISC of the tripdoublet to the lowest quartet state (which is almost always lower than the lowest tripdoublet state owing to Hund's rule). Both can be inhibited by making the molecule rigid, which is indeed satisfied by the porphyrin ligand in CuP. Alternatively, if the ISC from the first quartet state to the ground state is slow (as is the case of CuP, thanks to the spatial symmetry selection rules), and the gap between the first doublet and the first quartet is comparable to the thermal energy $kT$ at the current temperature, then the quartet state can undergo RISC to regenerate the tripdoublet state, which can then fluoresce. This is well-known as the thermally activated delayed fluorescence (TADF) mechanism\cite{TADF_old,TADF_Adachi,TADF_review}, although existing TADF molecules typically fluoresce from singlets and use a triplet ``reservoir state'' to achieve delayed fluorescence.
In order for the TADF mechanism to outcompete the phosphorescence from the first quartet state, both the phosphorescence rate and the doublet-quartet gap have to be small. While the low phosphorescence rate of CuP can be explained by the fact that copper is a relatively light element, the small $^2$T$_1$-$^4$T$_1$ gap of CuP can be attributed to the distributions of the frontier orbitals of CuP. Recall that the X-TDDFT gap between a tripdoublet excitation Eq.~\ref{Psi_tripdoublet} and the associated quartet excitation Eq.~\ref{Psi_quartet} is exactly given by the X-RPA gap\cite{XTDDFT}, which is equal to $\frac{3}{2}\left((it|it) + (ta|ta)\right)$. However, both of the two integrals are small for the $^2$T$_1$ and $^4$T$_1$ states of CuP, since the orbitals $i$ and $a$ reside on the ligand while $t$ is localized near the metal atom (Figure~\ref{CuP_orbitals}). Such a clean spatial separation of the metal and ligand CMOs (despite the close proximity of the metal and the ligand) can further be attributed to the fact that the Cu 3d$_{x^2-y^2}$ orbital has a different irrep than those of the ligand $\pi$ and $\pi^*$ orbitals, preventing the delocalization of the open-shell orbital to the $\pi$ system of the porphyrin ligand; while the Cu 3d$_{x^2-y^2}$ orbital can still delocalize through the $\sigma$ bonds of the ligand, the delocalization is of limited extent due to the rather local electronic structures of typical $\sigma$ bonds (Figure~\ref{CuP_orbitals}). Incidentally, the only other class of tripdoublet-fluorescing metalloporphyrins that we are aware of, i.e.~vanadium(IV) oxo porphyrin complexes\cite{Gouterman-Cu-VO-Co-porphyrins-1969,Gouterman-Cu-VO-porphyrins-1970}, are characterized by a single unpaired electron in the 3d$_{xy}$ orbital, whose mixing with the ligand $\pi$ and $\pi^*$ orbitals is also hindered by symmetry mismatches. Whether this can be extended to a general strategy of designing molecules that fluoresce from tripdoublet states (or more generally, molecules that possess small doublet-quartet gaps) will be explored in the future. Finally, we briefly note that the design of doublet molecules with TADF and/or phosphorescence is also an interesting subject and deserves attention in its own right.
\\
\section{Conclusion}
Fluorescence of open-shell molecules from tripdoublet states is a rare and underexplored phenomenon, for which traditional excited state methods such as U-TDDFT are unreliable due to severe spin contamination. In this work, we employed the high-precision method SDSPT2 to obtain accurate excitation energies of the CuP molecule, which suggests that the bright states obtained by light absorption relax to the lowest doublet state, $^2$T$_1$, via a cascade of ultrafast IC processes, in agreement with experiments. Contrary to predictions from ROKS orbital energy differences, $^2$T$_1$ is a tripdoublet state composed of a triplet ligand state antiferromagnetically coupled with the unpaired electron of Cu(II). Using the SDSPT2 results as a benchmark, we found that the X-TDDFT method provides a more accurate description of the $^2$T$_1$ state (which exhibits considerable spin contamination) compared to U-TDDFT, while for the CO excitations, U-TDDFT and X-TDDFT show similar performance.
\\\indent
In addition to vertical absorption calculations and structural analyses, we conducted a detailed analysis of the relaxation rate constants of the excited states of CuP. Our results suggest that, in the gas phase and at low temperature (83 K), CuP emits fluorescence from the lowest tripdoublet state $^2$T$_1$ with a very small quantum yield ($\sim 10^{-5}$), and the contribution of phosphorescence is negligible. These results complement the experimental results in solution phase and solid matrix, which gave a lower but still greater than unity fluorescence-to-phosphorescence ratio and a much higher luminescence quantum yield. Furthermore, we confirm the presence of an equilibrium between the first doublet state $^2$T$_1$ and the first quartet state $^4$T$_1$, the latter of which functions as a reservoir of the $^2$T$_1$ state, although the steady state concentration ratio of these two states deviates noticeably from their equilibrium constant. CuP therefore represents an interesting example of a TADF molecule that emits fluorescence through a doublet-doublet transition, instead of the much more common singlet-singlet pathway. Notably, U-TDA predicts a doublet-quartet gap of the wrong sign, due to the spin contamination of the doublet state as well as the breaking of the spin multiplet degeneracy of the quartet state. Although the error is small ($<$ 0.05 eV), it translates to a large error in the luminescence lifetime and (even more) the contribution of phosphorescence to the total lumincescence. This again highlights the importance of using spin-adapted approaches in the study of open-shell systems, even when the excitation energy errors of unrestricted methods are small.
\\\indent
Based on the computational results, we proposed a few possible approaches that can be used to design new doublet molecules that fluoresce from tripdoublets: (1) keep the open-shell orbital of the molecule spatially compact, to open up a gap between the $\alpha$ and $\beta$ UKS orbital energies of the open-shell orbital; (2) make the gap between the highest doubly occupied orbital and the lowest vacant orbital small enough so that both orbitals fit into the gap between the $\alpha$ and $\beta$ open-shell orbitals, but not overly small as to encourage IC of the lowest tripdoublet state to the ground state; (3) make the molecule rigid to minimize unwanted non-radiative relaxation processes; (4) avoid introducing heavy elements in order to suppress unwanted ISC and phosphorescence processes; and (5) localize the open-shell orbital and the frontier $\pi/\pi^*$ orbitals onto different molecular fragments, and (if possible) make them belong to different irreps, to minimize the doublet-quartet gap. We hope that the present work will facilitate the discovery of novel molecules that fluoresce from tripdoublet states. Moreover, we expect that the success of the X-TDDFT and SDSPT2 methods will encourage the use of these two methods in the excited state studies of other systems.

\section*{Conflict of Interest Statement}

The authors declare that the research was conducted in the absence of any commercial or financial relationships that could be construed as a potential conflict of interest.

\section*{Author Contributions}

WL and ZW conceived the topic, supervised the work and critically revised the manuscript. XW and ZW performed computational investigations and wrote the first draft. ZW and CW provided scientific advice and validated the data. All authors listed have made a substantial, direct, and intellectual contribution to the work and approved it for publication.

\section*{Funding}

This work was supported by the National Natural Science Foundation of China (Grant Nos. 21833001, 21973054, 22101155), Mountain Tai Climbing Program of Shandong Province, and Key-Area Research and Development Program of Guangdong Province (Grant No. 2020B0101350001). ZW gratefully acknowledges generous financial support by the Max Planck society.

\section*{Acknowledgments}

The authors acknowledge the computational software provided by the Institute of Scientific Computing Software in Shandong University and Qingdao BDF Software Technology Co., Ltd.

\section*{Data Availability Statement}

The original contributions presented in the study are included in the article/Supplementary Materials. Further inquiries can be directed to the corresponding authors.

\newpage


\bibliographystyle{apsrev4-1}
\bibliography{cup}


\end{document}